\documentclass[twocolumn,prd,nobibnotes,showpacs,showkeys,superscriptaddress,nofootinbib]{revtex4-1}
\usepackage[dvips]{graphicx,graphics,color,epsfig}
\usepackage[colorlinks=true,linkcolor=blue,citecolor=blue,urlcolor=blue,breaklinks=true]{hyperref}
\usepackage{natbib}
\usepackage{amsmath,amsfonts,amssymb}
\usepackage{latexsym}
\usepackage{dcolumn}
\usepackage{bm}
\begin{document}
\title
{\href{}
{Neutrino mass characteristics in a phenomenological $3+2+1$
model}
}
\author{N.~Yu.~Zysina}
\affiliation{National Research Center ``Kurchatov Institute'',
Kurchatov~place~1, 123182~Moscow, Russia}
\author{S.~V.~Fomichev}
\affiliation{National Research Center ``Kurchatov Institute'',
Kurchatov~place~1, 123182~Moscow, Russia}
\author{V.~V.~Khruschov}\email{khru@imp.kiae.ru}
\affiliation{National Research Center ``Kurchatov Institute'',
Kurchatov~place~1, 123182~Moscow, Russia}
\date{\today}
\begin{abstract}
The basic characteristics used for the description of both Dirac
and Majorana massive neutrinos are studied. Currently available
experimental data for these characteristics are presented with an
evidence of possible anomalies beyond the Standard Model with
three light neutrinos. Special attention has been paid to ways to
determine the neutrino mass absolute scale. In accordance with the
available data, the permissible values of the neutrino mass matrix
elements are found numerically against the minimal mass of
neutrino. Some phenomenological relations for the masses, the
angles and the {\it CP}-violating phases of the neutrino mixing
matrix are discussed, and the values of the neutrino mass
characteristics and the Dirac {\it CP}-violating phase are
evaluated for a model of bimodal neutrinos. The estimations made
for masses of sterile neutrinos, the values of the neutrino mass
characteristics and the Dirac {\it CP}-violating phase can be used
for interpretation and prediction of the results of various
neutrino experiments.
\end{abstract}

\maketitle %

\section{Introduction} %
\label{Section1} %
During the past years, a number of new interesting experimental
results were obtained in the neutrino physics. To such results,
one can refer the deviation from zero (more than $5\sigma$) of the
reactor mixing angle $\theta_{13}$ \cite{Fogli2012} and the
experimental indications on possible existing of new anomalies for
the number of neutrinos and antineutrinos in different processes
\cite{STV2011}. The first result is important for determination of
both the neutrino mixing matrix and the neutrino mass matrix
\cite{GGN2003}, and also for justification of the experimental
search of {\it CP} violation in the lepton sector \cite{BGFJ2012}.
As is known, the experimental data confirming lepton CP violation
are absent to the present. The second result can be connected with
existence of light sterile neutrinos, to which one can refer, for
example, SU(2)$_L$ singlet right neutrinos. The existence of right
neutrinos is beyond the Standard Model (SM) of weak and
electromagnetic interactions. However, the discovery of
oscillations of atmospheric, solar, reactor and accelerator
neutrinos, as well as extremely smallness of the neutrino mass
definitely points to violation of the conservation laws of the
leptonic numbers $L_{e}$, $L_{\mu }$ and $L_{\tau }$ and existence
of right neutrinos.

One of the unsettled outstanding problem in the neutrino physics
is the question about the neutrino nature, that is the question,
does neutrino belong to the Dirac or the Majorana particle type?
These two types of elementary particles were introduced in the
particle physics in 1928 by Dirac \cite{Dirac} and in 1937 by
Majorana \cite{Majorana}, respectively. In the present,
considerable efforts by both experimentalists and theorists are
concentrated to solve this problem. In this, it can be unexpected
the lack of unambiguous answer. For example, neutrinos can
simultaneously posses by both the Dirac and the Majorana
properties \cite{Wolfenstein1981,Petcov1982,Doi1983,Valle1983}.
One of the models, in which such situation is realized, is the
model of bimodal ({\it schizophrenic}) neutrinos
\cite{ADM2011,MP2011,CL2011}.

The other question requiring its speedy solution is the question
about the absolute scale of neutrino masses. As is known, from the
oscillation experiments with neutrinos it is possible to determine
the absolute values of the differences of the neutrino mass in
squares, rather than the mass values themselves. This circumstance
gives rise to the problem of the neutrino mass hierarchy, as well
as to the problem of the absolute mass scale. In spite of these
problems can be distinctly solved only with using the results of
the future special experiments, the different variants of their
solutions in the framework of the phenomenological models are also
proposed \cite{MS2006,Fritzsch2009,Gaponov2011,Khruschov2011}.

In the current paper, a phenomenological $3+2+1$ neutrino model is
proposed for solution of the problems in the neutrino physics
noted above. This model involves three active light neutrinos and
also three sterile neutrinos, from which one sterile neutrino is
comparatively heavy, but two other are light sterile neutrinos
\cite{Abazajian2012}. It is important that light sterile neutrinos
may be practically degenerate by mass with light active neutrinos,
so that they can jointly form quasi Dirac neutrinos, thus
realizing the bimodal neutrino states with two Majorana and two
quasi Dirac neutrinos. On the basis of available experimental
data, the acceptable ranges for neutrino mass characteristics in
this model are found numerically. Allowing for the cosmological
restrictions on total number of neutrinos \cite{Komatsu2011}, it
is easy to reduce the number of sterile neutrinos in the framework
of the model under consideration, that is to pass from the $3+2+1$
model to the $3+1+1$, or even the $3+1$ model. However, such a
reduction should be caused by forcible reasons. In this
connection, note that the experimentally obtained cosmological
restrictions on total number of neutrinos in turn are
model-dependent, and additional studies of the models permitting
existence of five or even six Majorana neutrinos, as well as more
accurate experimental restrictions on the number of neutrinos are
indispensable.

The paper is organized as follows. In Sec.~\ref{Section2}, the
basic characteristics used for the description of both the Dirac
and the Majorana massive neutrinos, as well as available
experimental data for these characteristics are present. For the
latter, the possible anomalies beyond the SM with three light
active neutrinos (with masses less than $m_{Z}/2$, where $m_{Z}$
is the mass of the $Z$--boson) are indicated. In
Sec.~\ref{Section3}, a special attention is paid to ways of
determination of the neutrino mass absolute scale, and also to
restrictions on this scale following from the experimental data.
In Secs.~\ref{Section4} and \ref{Section5}, respectively, some
phenomenological relations for the neutrino masses, as well as for
the mixing angles and the {\it CP}-violating phases of the
neutrino mixing matrix are discussed, and the values of the
neutrino mass observables and the Dirac {\it CP}-violating phase
for the $3+2+1$ neutrino model are calculated. The results of the
paper, which can be used for the interpretation and prediction of
the results of various neutrino experiments are discussed in
Sec.~\ref{Section6}, which terminates the paper.

\section{Experimental values of the basic characteristics
of the Dirac and Majorana massive neutrinos}
\label{Section2} %
It is known that the oscillations of the solar, atmospheric,
reactor and accelerator neutrinos can be explained by mixing the
neutrino states. It means that the flavor states of neutrino are
the mix, at least, of three massive neutrino states, and vice
versa. Neutrino mixing is described by the
Pontecorvo--Maki--Nakagawa--Sakata matrix $U_{P\!M\!N\!S}\equiv
U=VP$, which enters the relation
\begin{equation}
\psi_{\alpha L}=U_{\alpha i}\psi_{iL},\label{eq1}
\end{equation}
where $\psi_{\alpha L}$ and $\psi_{iL}$ are the left chiral flavor
or massive neutrino fields, respectively, with
$\alpha=\{e,\mu,\tau\}$ and $i=\{1,2,3\}$, and summation over
repeated indices is implied. The matrix $V$ can be written in the
standard parameterization \cite{Nakamura2010} as
\begin{widetext}
\begin{equation}
V=\left(\begin{array}{*{20}c}
c_{12}c_{13}\hfill & s_{12}c_{13}\hfill & s_{13}e^{-i\delta}\hfill\\
-s_{12}c_{23}-c_{12}s_{23}s_{13}e^{i\delta}\hfill & c_{12}
c_{23}-s_{12}s_{23}s_{13}e^{i\delta}\hfill & c_{13}s_{23}\hfill\\
s_{12}s_{23}-c_{12}c_{23}s_{13}e^{i\delta}\hfill & -c_{12}
s_{23}-s_{12}c_{23}s_{13}e^{i\delta}\hfill & c_{13}c_{23}\hfill\\
\end{array}\right),
\label{eq2}
\end{equation}
\end{widetext}
through the quantities $c_{ij}\equiv\cos\theta_{ij}$ and
$s_{ij}\equiv\sin\theta_{ij}$, and with the phase $\delta$
associated with the Dirac {\it CP} violation in the lepton sector.
On the other hand, the $3\times3$ matrix $P$ is the diagonal one,
$P={\rm diag}\left\{1,e^{i\alpha},e^{i\beta}\right\}$, with
$\alpha$ and $\beta$ the phases associated with the Majorana {\it
CP} violation.

Generally, an unitary $n\times n$ matrix is defined by $n^{2}$
real parameters, which can be chosen as $n(n-1)/2$ angles and
$n(n+1)/2$ phases. Taking into account the structure of the SM
electro-weak lagrangian involving currents of quarks, charged
leptons and neutrinos, in the case of the Dirac neutrinos it is
possible to exclude $2n-1$ phases from these parameters. In the
opposite case, if all the neutrinos are the Majorana type
particles, it is possible to exclude only $n$ phases associated
with the Dirac charge leptons. Thus, depending on the neutrino
type, the $n\times n$ $U_{P\!M\!N\!S}$ matrix is defined by either
$n(n-1)/2$ angles and $(n-1)(n-2)/2$ phases, if neutrinos belong
to the Dirac type, or $n(n-1)/2$ angles and $n(n-1)/2$ phases, if
neutrinos are the Majorana-type particles \cite{BGG1999}.

With the help of the neutrino mixing matrix $U$, the mass matrix
$M$ of three active neutrinos can be determined as follows
\begin{equation}
M=U^{\ast}M_{d}U^{+}\,,
\label{eq3}
\end{equation}
where $M_d={\rm diag}(m_1,m_2,m_3)$, with $m_i$ the neutrino
masses. Matrix elements $M_{ij}$ of the matrix $M$ depend on the
neutrino masses and the mixing parameters, that is the mixing
angles and phases. For the case of three light Dirac neutrinos, in
addition to masses it is necessary to determine three mixing
angles ($\theta_{12}$, $\theta_{13}$ and $\theta_{23}$) and one
mixing phase ({\it CP}-violating phase $\delta$), while for three
Majorana neutrinos it requires to determine three mixing angles
($\theta_{12}$, $\theta_{13}$ and $\theta_{23}$) and three {\it
CP}-violating phases ($\delta$, $\alpha$ and $\beta$).

The determination of the neutrino mass absolute values and the
mixing parameters, as well as ascertainment of the neutrino type
(Dirac vs Majorana) are at present the basic problems of the
neutrino physics. Solutions of these problems require active
experimental and theoretical studies of both the neutrino mass
observables, which define the absolute mass scale, and the
neutrino oscillation characteristics, which characterize mixing of
the neutrino states with different masses. It is expected that the
comprehensive solution of the problem of the neutrino nature, as
well as of the neutrino masses and mixing parameters will be given
in the future Grand Unification Theory (GUT), which does not exist
at present in its conventional form \cite{Nakamura2010}. A
required direction of the further development of the SM can also
be found by study of different correlations between experimental
values of the neutrino masses and mixing parameters, to which we
refer the angles of mixing and the {\it CP}-violating phases. The
discovery of such interrelationships can play an important role in
finding out the ways to expand the SM and to successfully develop
at last the following consistent GUT. For this, the data received
in the current and the future experiments on determination the
neutrino characteristics (PLANK, KATRIN, GERDA, CUORE, BOREXINO,
Double CHOOZ, SuperNEMO, KamLand-Zen, EXO, etc.) will undoubtedly
play a decisive role.

With using the only oscillation experiments with atmospheric,
solar, reactor and accelerator neutrinos it is impossible to
determine the neutrino mass absolute values, as well as the type,
either Majorana or Dirac, of the neutrino. However, the
experimental data obtained in the neutrino oscillation experiments
indicate the violation of the conservation laws of the leptonic
numbers $L_{e}$, $L_{\mu}$, $L_{\tau}$, and, besides, by virtue of
deviation from zero of two oscillation parameters $\Delta
m^{2}_{12}$ and $\Delta m^{2}_{13}$ (with $\Delta m_{ij}^2=m_i^2-
m_j^2)$ they indicate the existence at least two nonzero and
different neutrino masses. Below we present the experimental
values of the mixing angles and the oscillation parameters, which
determine three-flavor oscillations of the light neutrinos.
Together with the standard uncertainties on the level of
$1\sigma$, these data obtained as a result of a global analysis of
the latest high-accuracy measurements of the oscillation
parameters \cite{Fogli2012} are as follows
\begin{subequations}
\begin{align}
&\sin^2\theta_{12}=0.307_{-0.016}^{+0.018}\,,\label{eq4a}\\
&\sin^2\theta_{23}=\left\{\begin{array}{*{20}c}
NH:\hfill & 0.386_{-0.021}^{+0.024}\hfill\\
IH:\hfill & 0.392_{-0.022}^{+0.039}\hfill\label{eq4b}\\
\end{array}\right.,\\
&\sin^2\theta_{13}=\left\{\begin{array}{*{20}c}
NH:\hfill & 0.0241_{-0.0025}^{+0.0025}\hfill\\
IH:\hfill & 0.0244_{-0.0025}^{+0.0023}\hfill\label{eq4c}\\
\end{array}\right.,\\
&\Delta m_{21}^2/10^{-5}{\rm
eV}^2=7.54_{-0.22}^{+0.26}\,,\label{eq4d}\\
&\Delta m_{31}^2/10^{-3}{\rm eV}^2=\left\{\begin{array}{*{20}c}
NH:\hfill & 2.47_{-0.10}^{+0.06}\hfill\\
IH:\hfill & -2.46_{-0.11}^{+0.07}\hfill\label{eq4e}\\
\end{array}\right..
\end{align}
\label{eq4}
\end{subequations}
Since only the absolute value of $\Delta m_{31}^2$ is known, it is
possible to arrange the absolute values of the neutrino masses by
two ways, namely, as
\begin{equation}
a)\,\,m_1<m_2<m_3\quad{\rm and}\quad b)\,\,m_3< m_1<m_2\,.
\label{eq5}
\end{equation}
These two cases correspond to so called the normal hierarchy (NH)
and the inverse hierarchy (IH) of the neutrino mass spectrum,
respectively. Unfortunately, the $CP$-violating phases $\alpha$,
$\beta$ and $\delta$, as well as the neutrino mass absolute scale
are unknown at present.

Three groups of the experimental data associated with the neutrino
are sensitive to the neutrino absolute mass scale, namely, they
are the data on $\beta$ decay, the data on neutrinoless
double-beta decay, and the data obtained as a result of the
cosmological observations. So, to determine the neutrino absolute
mass scale it is necessarily to determine experimentally at least
one from three mass observables of the neutrino, namely, either
the mean cosmological mass $m_{a}$ of the active neutrinos, or the
$\beta$ decay neutrino mass $m_{\beta}$, or the effective
double-beta decay neutrino mass $m_{\beta\beta}$, which are
defined as follows
\begin{subequations}
\begin{align}
&m_a=\frac{1}{3}\sum\nolimits_{i\,=\,1,2,3}{\vert
m_i\vert}\,,\label{eq6a}\\
&m_\beta=\Big(\sum\nolimits_{i\,=\,1,2,3}{\vert U_{ei}\vert}
^2m_i^2\Big)^{1/2},\label{6b}\\
&m_{\beta\beta}=\Big|\sum\nolimits_{i\,=\,1,2,3}U_{e{\kern0.5pt}i}^2\,m_i
\Big|\,,\label{eq6c}
\end{align}
\label{eq6}
\end{subequations}
where $U_{e{\kern0.25pt}i}$ are the elements of the
Pontecorvo--Maki--Nakagawa--Sakata matrix. Generally, we will call
both the neutrino mass observables and the neutrino masses as the
mass characteristics of the neutrino. With help of the
corresponding data from these three groups of the experiments
indicated above, three mass observables $m_{a}$, $m_{\beta}$, and
$m_{\beta\beta}$ one by one can be determined, respectively.
Currently, the upper limits are only obtained for the mass
observables, namely, $m_{a}<0.2$~eV \cite{Komatsu2011},
$m_{\beta}<2.2$~eV \cite{OW2008}, and $m_{\beta\beta}<0.35$~eV
\cite{KK2001,Aalseth2004}, where in the last case the limit should
be increased approximately one and a half or even twice to take
into account the uncertainties in the nuclear matrix elements
governing this limitation. Note that the limit for $m_{\beta}$
given above, which was obtained in the experiments carried out in
Troitsk and Mainz on the electron spectrum measurements in the
tritium $\beta$ decay, is planned to be improved up to $0.2$~eV in
the scheduled KATRIN experiment \cite{OW2008}.

As we know, the right neutrinos are sterile particles and are the
candidates, along with the other possibilities, to the specific
particles of the dark matter. There can be several types of the
dark matter particles. We consider only the right neutrinos
$\nu_{R}$. If $\nu_{R}$ together with the $\nu_{L}$ form the Dirac
mass term in the Lagrangian, then $\nu_{R}$ will be degenerate in
mass with $\nu_{L}$ thus being the light particles with the mass
less than, at least, $0.3$~eV. The other possibility is that the
masses of $\nu_{R}$ can be larger than the masses of $\nu_{L}$.
For example, they can belong to the range from $0.3$~eV up to
$3$~eV. The right neutrinos of such a kind can be called as heavy
sterile neutrinos. Besides, the neutrinos with masses from $3$~eV
up to $3$~GeV and with masses more than $3$~GeV can be called as
extra-heavy and super-heavy neutrinos, respectively. The existence
of super-heavy right neutrinos can be used both for explanation of
high value of the invisible mass of the Universe, and for the
explanation of small masses of the left neutrinos. Moreover, with
the super-heavy right neutrinos, the observed baryon asymmetry of
the Universe can be explained \cite{FY1986}.

Recently \cite{Mueller2011}, the corrected calculations of the
spectrum of reactor antineutrino were provided, which result in
higher calculated values of the fluxes of these particles. Thus,
the experimental data indicate on the antineutrino deficiency in
the measurements of the antineutrino fluxes on distances lower
than $100$~m from the particle source. These distances from the
source should be considered as small. The currently available
indications on the antineutrino deficiency on small distances can
result in reactor anomaly. Rather like anomalies were observed in
calibration measurements for the experiments GALLEX and SAGE
\cite{Abdurashitov2006,GL2010}. These anomalies can be called as
calibration ones. The most recent experimental data indicating
that the angle $\theta_{13}$ is noticeably deviated from zero [cf.
Eq.~(\ref{eq4c})] enhance the probability of the reactor and
calibration anomalies. In this, the estimation
$\sin^2\theta_{14}\lesssim0.04$ for the mixing angle $\theta_{14}$
of the active and sterile neutrinos is obtained in the $3_a+1_s$
model with single sterile neutrino
\cite{Abazajian2012,Palazzo2012}. Note that, besides the $3_a+1_s$
model, the $3_a+2_s$ model with three active and two sterile
neutrinos is used for explanation of the observed anomalies of the
spectra of neutrinos and antineutrinos, including the neutrino
spectra anomalies in the LSND experiment \cite{Aguilar2001}, and
then in the MiniBooNE experiment \cite{AguilarArevalo2010} and in
the calibration measurements \cite{Abdurashitov2006,GL2010}. In
these models, the typical values of $\Delta m^2$ are of the order
of $1$ eV$^2$.

\section{Evaluation of the neutrino mass absolute scale from the possible
values of the neutrino mass observables}
\label{Section3} %
\begin{figure}
\centering
\includegraphics[width=0.49\textwidth]{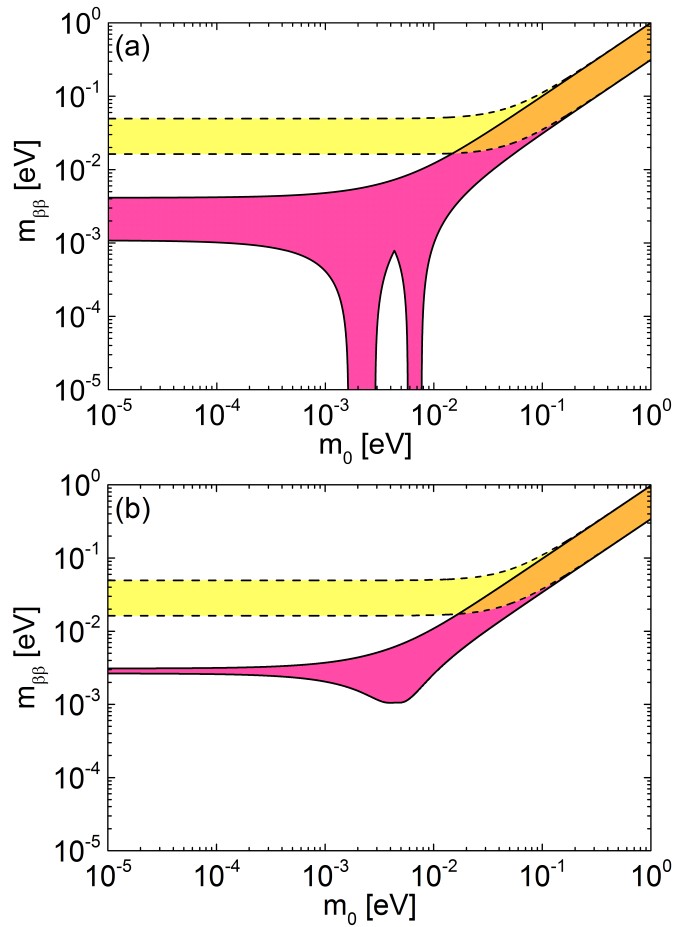}
\caption{(Color online) The variation ranges of the effective
neutrino mass $m_{\beta\beta}$ as a function of the minimal
neutrino mass $m_0$ at $\delta=0$ (a) and at $\delta=45$\!\r{}
(b), at $\alpha=0$, $\beta=0$ and at the values of the oscillation
parameters from Eqs.~(\ref{eq4}). The shaded areas
restricted by the solid and the dashed curves
correspond to the NH and IH cases, respectively.}\label{Fig:1}
\end{figure}
\begin{figure}
\centering
\includegraphics[width=0.45\textwidth]{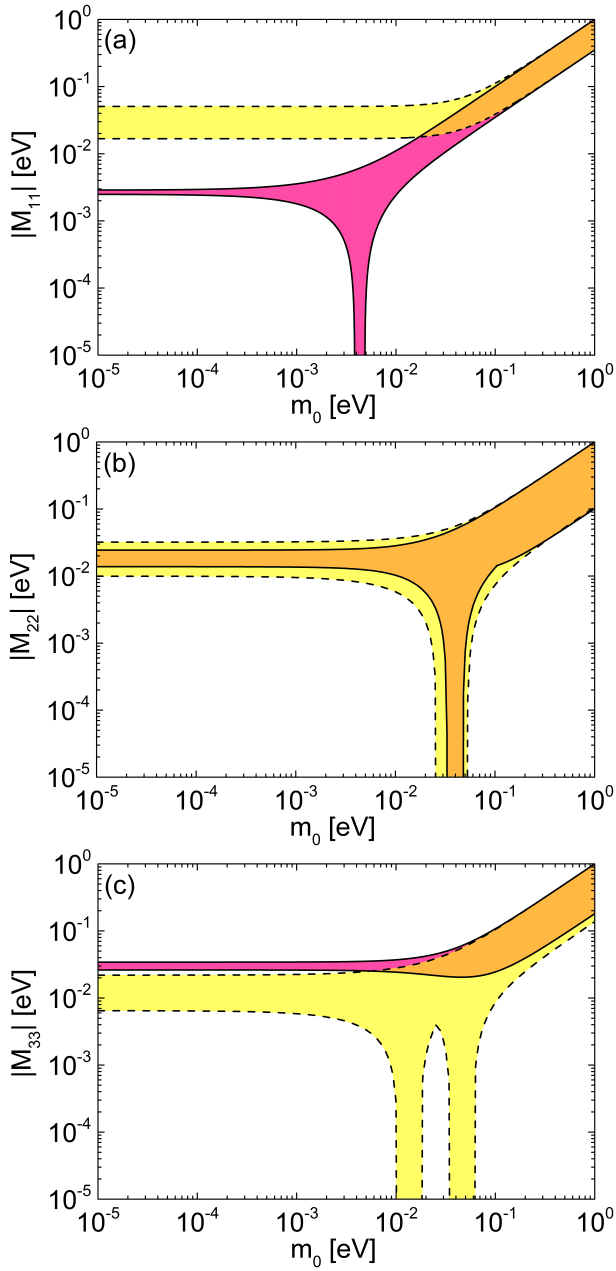}
\caption{(Color online) The ranges of absolute values of the matrix elements
$M_{11}$ (a), $M_{22}$ (b) and $M_{33}$ (c) versus the minimal neutrino mass 
$m_0$ at $\sin^2\theta_{13}=0$, $\delta=0$, $\alpha=0$, $\beta=0$. 
The values of the other oscillation parameters coincide with the data 
from Eqs.~(\ref{eq4}). The shaded areas restricted by the solid and the dashed 
curves correspond to the NH and IH cases, respectively.}\label{Fig:2}
\end{figure}
\begin{figure}
\centering
\includegraphics[width=0.45\textwidth]{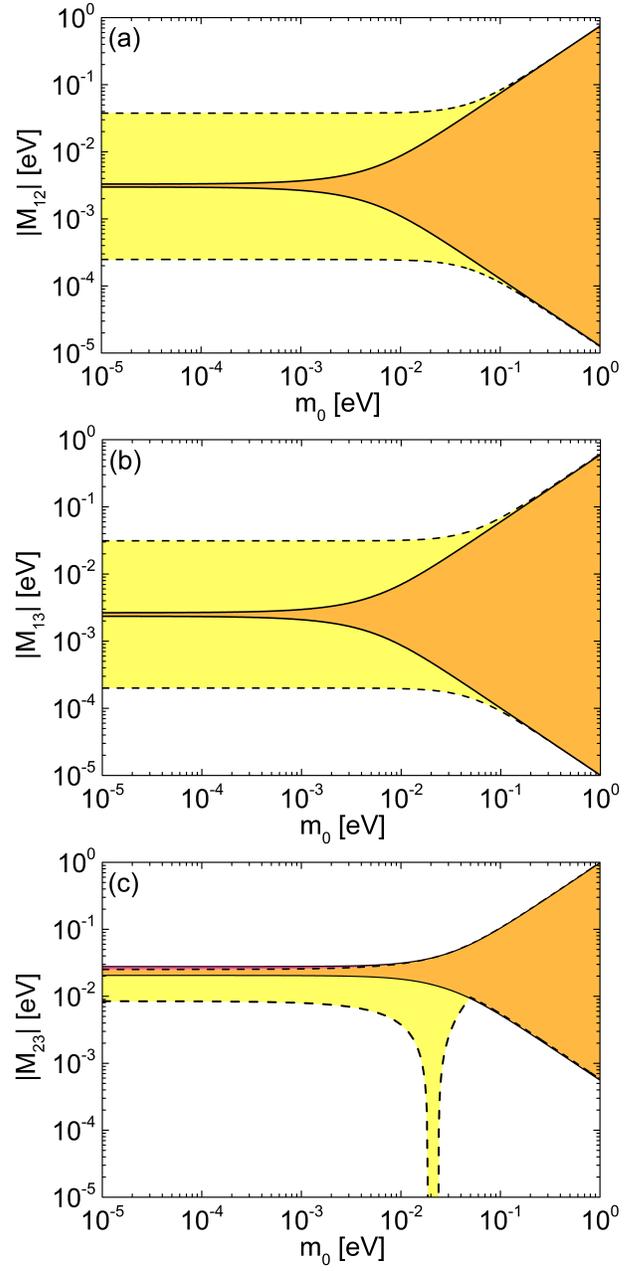}
\caption{(Color online) The same as in Fig.~\ref{Fig:2}, but
for off-diagonal matrix elements $M_{12}$ (a), $M_{13}$ (b)
and $M_{23}$ (c) of the neutrino mass matrix.
}\label{Fig:3}
\end{figure}

Let us consider the neutrino mass matrix given by Eq.~(\ref{eq3}).
The upper diagonal matrix element $M_{11}$ of this matrix enters
the relation for the probability of the nuclear neutrinoless
double-beta decay $(A,Z)\to(A,Z+2)+2e$, if the decay occurs with
the assistance of the Majorana light neutrinos. In this case the
absolute value of $M_{11}$ coincides with $m_{\beta\beta}$ from
Eq.~(\ref{eq6c}), i.e., with the neutrino effective mass. The
nuclear half-life period $T_{1\!/2}^{0\nu\,2\beta}$ is inversely
proportional to $m_{\beta\beta}^2$ \cite{BG2012}. Note that the
discovery of the neutrinoless double-beta decay is practically the
only way to determine whether neutrinos are of the Dirac-type or
the Majorana-type particles. The discovery of this decay could
make it possible to determine the neutrino absolute mass scale
with the aid of the $m_{\beta\beta}$ experimental value.

The explicit expression for $m_{\beta\beta}$ through the neutrino
masses and the matrix elements of the neutrino mixing matrix
$U_{P\!M\!N\!S}$ is given in Eq.~(\ref{eq6c}). In the present, on
the basis of the experimental data given in Eqs.~(\ref{eq4}) on
the mixing parameters of neutrino oscillations, and with the
adjusting values of the $m_{\beta\beta}$, it can be possible to
estimate the absolute scale of the neutrino mass spectra with the
normal and inverse hierarchy. Indeed, $m_{\beta\beta}$ is
expressed trough the neutrino masses and the neutrino mixing
parameters as follows
\begin{equation}
m_{\beta\beta}=|c_{12}^2c_{13}^2m_1+s_{12}^2c_{13}^2e^{2i\alpha}m_2
+s_{13}^2e^{2i(\beta-\delta)}m_3|\,.\label{eq7}
\end{equation}
With using Eq.~(\ref{eq7}) and the experimental data from
Eqs.~(\ref{eq4}), the explicit dependences of $m_{\beta\beta}$
versus the minimal neutrino mass $m_{0}$, that is either versus
$m_{1}$ in the NH case or versus $m_{3}$ in the IH case can be
determined. Besides, the similar dependences can be found for
absolute values of other matrix elements $M_{ij}$.
Figure~\ref{Fig:1}\textcolor{blue}{(a)} exhibits the dependences of $m_{\beta\beta}$
versus $m_{0}$ jointly for both NH and IH cases at zero value of
the neutrino {\it CP}-violating phases. The allowance for the
nonzero {\it CP}-violating phases alters the valid (from the point
of view of available experimental data) range of variation of
$m_{\beta\beta}$ versus $m_{0}$. For example, the characteristic
limitation of the $m_{\beta\beta}$ values from the bottom arises
at some values of the {\it CP}-violating phase $\delta$. Such
behavior of $m_{\beta\beta}$ for both NH and IH cases and for
$\delta=45$\!\r{} is presented in Fig.~\ref{Fig:1}\textcolor{blue}{(b)}.

It is interesting that increase of accuracy of the numerical
calculations together with the new experimental data results in
division of the range of the minimal neutrino mass in the NH case
on two sub-ranges, where zeroth values of $m_{\beta\beta}$ can be
reached. At values of $m_{0}$ between these two sub-ranges,
$m_{\beta\beta}$ is restricted from the bottom. These new features
of behavior of $m_{\beta\beta}$ as a function of $m_{0}$ were not
noted previously.

The possible structure of the neutrino mass matrix, as well as
applicability of some additional assumptions about values of its
matrix elements, which are used in a number of phenomenological
models \cite{Fritzsch2011,MB2012}, can be obtained numerically.
For calculations of the absolute values of the matrix elements
$M_{ij}$, the approximation of $s_{13}=0$ can be safely used.
Figures~\ref{Fig:2} and \ref{Fig:3} exhibit the absolute values of
diagonal and off-diagonal elements $M_{ij}$, respectively, versus
the minimal neutrino mass $m_0$ for both NH and IH cases. Note
that in the applied approximation of $s_{13}=0$, the main features
of behavior of $|M_{ij}|$ remain invariable, but the fine
peculiarities such a division of the $m_0$-range, where
$m_{\beta\beta}$ vanishes, on the two sub-ranges disappear (cf.
Fig.~\ref{Fig:1}\textcolor{blue}{(a)} and Fig.~\ref{Fig:2}\textcolor{blue}{(a)} for $|M_{11}|$). These
dependences permit us to conceive of the structure of mass matrix
$M_{ij}$ at different values of $m_{0}$. For example, at
$m_{0}\approx3$~meV (the mean values of $M_{ij}$ are also
presented in~meV)
\begin{equation}
M^{(NH)}=\begin{pmatrix}
2 & 3 & 2\\
3 & 15 & 14\\
2 & 14 & 12
\end{pmatrix},\quad
M^{(IH)}=\begin{pmatrix}
30 & 3 & 2\\
3 & 20 & 10\\
2 & 10 & 20
\end{pmatrix}.
\label{eq8}
\end{equation}
The different structures of matrix $M_{ij}$ and different
assumptions about the values of its matrix elements were
considered in different models
\cite{Fritzsch2011,MB2012,MR2006,CPP2006}. For instance, the
conditions of vanishing of the individual matrix elements, as well
as of its spur and determinant were analyzed. Figures~\ref{Fig:2}
and \ref{Fig:3} for the absolute values of the matrix elements
$M_{ij}$ can be used for inspection of different models of the
neutrino mass matrix structure.

As can be seen from Fig.~\ref{Fig:1}, while specifying the
permissible values of $m_{\beta\beta}$ it can be possible to
determine the eventual intervals of the values of the minimal
neutrino mass $m_{0}$, and then the absolute values of the other
neutrino masses. Note that such calculations of the neutrino mass
absolute values are sufficiently precise at small values of
$m_{\beta\beta}$, in spite of uncertainties of the experimental
data. They are consistent with the results of the papers
\cite{PP2004,PPS2006} in the limits of the precision of the
presented graphs and the data used, in which the conditions
resulting in the values of $m_{\beta\beta}$ greater than
$10^{-3}$~eV were considered. As is seen from Fig.~\ref{Fig:1}\textcolor{blue}{(a)},
the IH spectrum can not be realized for $\delta=0$ at
$m_{\beta\beta}<0.01$~eV that was noted before in
Ref.~\cite{PP2004}. Besides, as a result of the latter work, there
is the restriction of the NH spectrum at small values of
$m_{\beta\beta}$ at some values of $\delta$ [cf. with
Fig.~\ref{Fig:1}\textcolor{blue}{(b)}], as well as at $\delta=0$ in a small
intermediate range [cf. with Fig.~\ref{Fig:1}\textcolor{blue}{(a)}].

\begin{figure}
\centering
\includegraphics[width=0.49\textwidth]{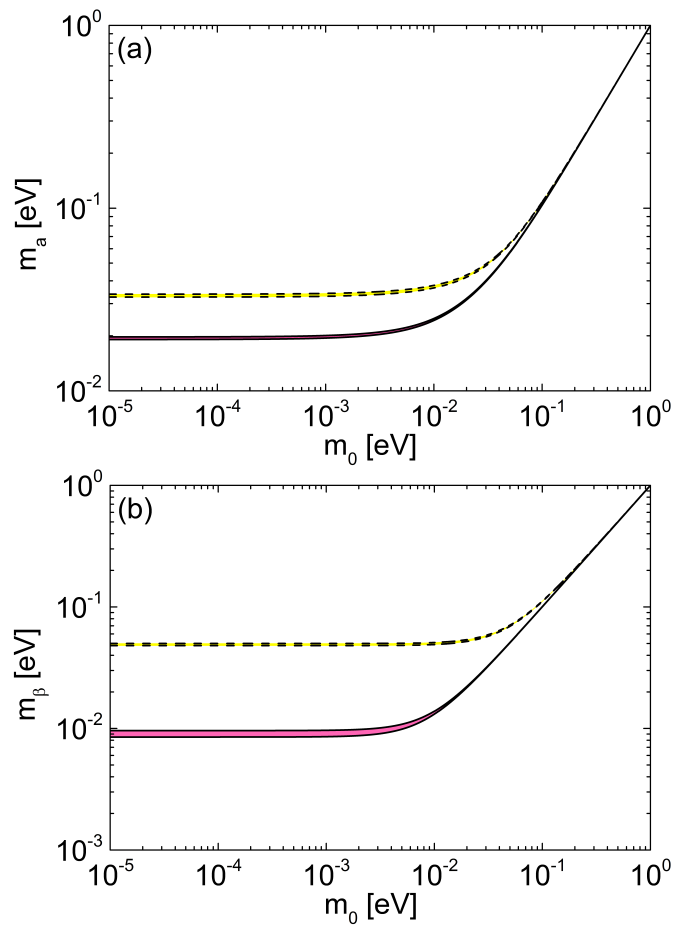}
\caption{(Color online) The mean cosmological mass of active
neutrinos $m_a$ (a) and the $\beta$ decay neutrino mass
$m_{\beta}$ (b) at the oscillation parameters from
Eqs.~(\ref{eq4}). On both panels, the lower and upper narrow ranges
restricted by the solid and the dashed curves
correspond to the NH and IH cases, respectively.} \label{Fig:4}
\end{figure}

In Figures~\ref{Fig:4}\textcolor{blue}{(a)} and \textcolor{blue}{(b)}
we present the mean cosmological
mass $m_{a}$ of active neutrinos and the $\beta$ decay neutrino
mass $m_{\beta}$, respectively, versus the minimal neutrino mass
$m_0$, which were obtained with the latest experimental data from
Eqs.~(\ref{eq4}) on the oscillation characteristics of the
neutrino.

The obtained intervals of the possible values of $m_{\beta\beta}$
can be used for planning the experiments on search of the
neutrinoless nuclear double-beta decay and for interpretation of
the results obtained with allowance for the {\it CP} violation.
The neutrino mass observables $m_{a}$ and $m_{\beta}$ can be used
for the experiments with the results dependent on the neutrino
mass absolute values.

\section{Phenomenological relations for the neutrino masses}
\label{Section4} %
As is well-known, the problem of origin of the mass of the
fundamental fermions is still unsolved. In the SM, these masses
are originated due to Yukawa couplings between the fundamental
fermion fields and the Higgs fields. Since the actual properties
of the Higgs boson are still unknown in detail, the generation
mass mechanism accepted in the SM should be considered as a
hypothesis. Moreover, the values of the neutrino masses are so
small that, probably, the mechanism of the neutrino mass
generation is primarily connected with the possible Majorana
nature of the neutrino, rather than with the Higgs boson
properties. In this case, the immediate basic problem is the
simulation of the mass generation mechanism of the Majorana
neutrinos. In absence of the appropriate theory of this
phenomenon, the problem can be considered on the phenomenological
level \cite{Khruschov2011}. Let us suppose that there are several
different contributions to the mass of the neutrino, and two of
them are most significant. It can be assumed that the first
contribution results in the mass of the left light neutrino, which
is specified by the Majorana mass term in the Lagrangian as
\begin{equation}
L_m^{'}=-\bar{\nu}_LM_{L}\nu_L^c/2+{\rm h.c.}\label{eq9}
\end{equation}
In this case, with the Higgs mechanism of the mass generation, the
Higgs sector of the SM should be changed and expanded. In what
follows, the contribution of the type, which can be associated
with Eq.~(\ref{eq9}), will be taken into account with the help of
a new phenomenological parameter $\xi$. The second contribution
can be connected with the so called seesaw mechanism, which is
realized under inclusion into the theory the heavy right neutrinos
$N_i$, with $i=1,2,3$. This mass term is of the form
\begin{equation}
M_\nu^{''}=-M_D^T M_R^{-1}M_D\,,
\label{eq10}
\end{equation}
where $M_{D}$ is the matrix of the Dirac terms in the neutrino
masses, $M_{R}$ is the typical value of the right neutrino masses,
which establishes a new scale associated with the right neutrino
masses. Let us suppose that $M_{D}$ is proportional to the mass
matrix of the charged leptons, that is $M_{D}=\sigma M_l$, where
$\sigma$ is of the order of unity,
$M_l=diag\{m_e,m_{\mu},m_{\tau}\}$, and $M_{R}=\sigma^2M$. Then,
for estimations of the neutrino masses $m_i$, the following
phenomenological formula can be used \cite{Khruschov2011}:
\begin{equation}
m_{i}=\pm\xi-m_{li}^2/M, \label{eq11}
\end{equation}
with $m_{li}$ the masses of three charged leptons. With the help
of Eq.~(\ref{eq11}) and the data from Eqs.~(\ref{eq4}), it is
easily to obtain the absolute values of the neutrino masses $m_i$
and the typical scales of $\xi$ and $M$ in~eV for both the NH and
the IH cases. They are as follows, respectively,
\begin{subequations}
\begin{align}
&m_{1}\approx0.0693,\,m_{2}\approx0.0698,\,m_{3}
\approx0.0851,\nonumber\\
&\xi\approx0.0693,\,
M\approx2.0454\times10^{19},\,(NH),\label{eq12a}
\end{align}
\begin{align}
&m_{1}\approx0.0775,\,m_{2}\approx0.078,\,m_{3}\approx
0.0606,\nonumber\\
&\xi\approx0.0775,\,M\approx2.2872
\times10^{19},\,(IH).\label{eq12b}
\end{align}
\label{eq12}
\end{subequations}

In the mass neutrino generation scheme considered above, the right
neutrinos are super-heavy. However, the neutrino mass terms
defined by Eq.~(\ref{eq9}) and Eq.~(\ref{eq10}) can be given in a
different way, if to consider that one of the cause originating
the neutrino masses, both for the Majorana and the Dirac
neutrinos, is the interaction with the cosmological scalar field
of the order of the cosmological $\Lambda$-term. Then, both the
Majorana masses of the left neutrinos and the Dirac masses may be
of the same order with the typical linear mass scale $\lambda$ of
the cosmological $\Lambda$-term, it being known that the value of
$\lambda$ is equal approximately $2$~meV \cite{Komatsu2011}. In
this case, the mass absolute values of three light active
neutrinos $m_i$ will be equal in~eV for the NH and IH cases,
respectively, to
\begin{subequations}
\begin{align}
&m_{1}\approx0.002,\,m_{2}\approx0.0087,
\,m_{3}\approx 0.0497,\,(NH),\label{eq13a}\\
&m_{1}\approx0.0496,\,m_{2}\approx0.050,\,
m_{3}\approx0.002,\,(IH),\label{eq13b}
\end{align}
\label{eq13}
\end{subequations}
while the masses of the right neutrinos $M_i$ can be estimate as
follows
\begin{subequations}
\begin{align}
&M_{1}\approx\Lambda_1,\,M_{2}\approx0.002,\,M_{3}\approx
0.002,\,(NH),\label{eq14a}\\
&M_{1}\approx0.002,\,M_{2}\approx0.002,\,
M_{3}\approx\Lambda_3,\,(IH),\label{eq14b}
\end{align}
\label{eq14}
\end{subequations}
with $\Lambda_{1,3}$ the free parameters of the order of $1$~eV.
Note that in this case the possibility exists to identify the
right neutrinos $\nu_{Ri}$ with the sterile neutrinos $\nu_{si}$
($i=1,2,3$). This possibility leads to existence of three sterile
neutrinos, two of them are light and the third one is heavy. Let
us call this case with three light active neutrinos, one heavy
sterile neutrino and two light sterile neutrinos as the $3+2+1$
model. In case of need, this model can be reduced to the $3+1+1$,
or even to the $3+1$ model with the exception of the light right
neutrinos. However, it should be some additional weighty reasons
for absence of either one or even two right neutrinos. The light
right sterile neutrinos can be combined with light left neutrinos
to form the quasi Dirac neutrinos. The case of quasi Dirac
neutrinos is considered minutely in the next Sec.~\ref{Section5}.

\section{Phenomenological relations for the angles and
the {\it CP}-violating phases of the neutrino mixing matrix and
the model of bimodal neutrino}
\label{Section5} %
The upper diagonal matrix element $m_{\beta\beta}$ of the neutrino
mass matrix is connected with the probability of the neutrinoless
double-beta decay. At the same time, two other diagonal matrix
elements can be equal in absolute values if the $\mu-\tau$
symmetry is taken into account. It does not contradict to a number
of models of the neutrino mass matrix \cite{Fritzsch2011,MB2012}
and to the approximate $\mu-\tau$ symmetry \cite{GL2012}, as well
as to estimations obtained in Sec.~\ref{Section3} for the matrix
elements $M_{ij}$ [see. e.g., Eq.~(\ref{eq8})]. However, even in
this approximation it is impossible to obtain without additional
assumptions the estimations of $m_{\beta\beta}$ in the presence of
the {\it CP} violation\footnote{Without the {\it CP} violation,
the estimations for $m_{\beta\beta}$ can be easily obtained as
$0.07$ (NH) and $0.08$~eV (IH) at the values of neutrino masses
from Eqs.~(\ref{eq12}), while it will be equal to $0.005$ (NH) and
0.05~eV (IH) at the mass values from Eqs.~(\ref{eq13}).}. To
obtain such estimations, one can consider the case of bimodal
neutrino \cite{ADM2011,MP2011,CL2011,Khruschov2011}, when the
neutrino is neither the Dirac one nor the Majorana one but has
simultaneously both quasi Dirac and Majorana properties. Indeed, a
pair of quasi degenerate Majorana neutrinos can form the states of
quasi, pseudo, and imaginary Dirac particles. Quasi Dirac
particles consist of a pair of one active and one sterile
neutrino, pseudo Dirac particles consist of a pair of active
neutrinos, and imaginary Dirac particles consist of a pair of
sterile neutrinos. Since the investigations of such particles are
only starting now, the terminology given above is not completely
established. Such particles have the properties of the Dirac
particles and become entirely the Dirac particles under full
degeneration.

In the model of bimodal neutrino it is usually assumed that the
states of neutrino with a certain mass involve both the Majorana
and the quasi/pseudo Dirac particles. Let us consider the typical
case when, from three active neutrinos, one neutrino is purely
Majorana neutrino, while two other ones are quasi Dirac particles.
In the NH case, the second and the third neutrino mass states are
the quasi Dirac ones, while in the IH case they are
correspondingly the first and the second mass neutrino states.

\begin{figure}
\centering
\includegraphics[width=0.49\textwidth]{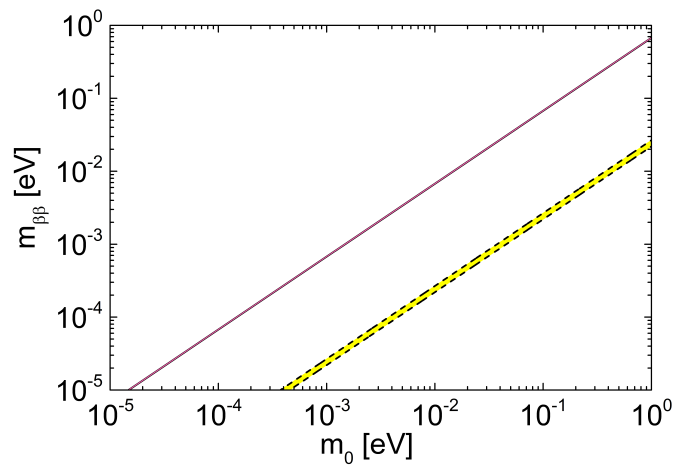}
\caption{(Color online) The $m_{\beta\beta}$ versus the minimal
neutrino mass $m_0$ in the model of bimodal neutrinos at the
oscillation parameters from Eqs.~(\ref{eq4}). The upper and lower
narrow ranges restricted by the solid and the dashed lines
correspond to the NH and IH cases, respectively.}
\label{Fig:5}
\end{figure}

Then, the condition of the $\mu-\tau$ symmetry, i.e., the equality
of $m_{\mu\mu}$ and $m_{ee}$ results in the relation, which in the
{\it NH} case reads as
\begin{align}
&|s_{12}^2c_{23}^2+2s_{12}c_{12}s_{23}c_{23}s_{13}e^{i\delta}
+c_{12}^2s_{23}^2s_{13}^2e^{2i\delta}|\nonumber\\
&=|s_{12}^2s_{23}^2-2s_{12}c_{12}s_{23}c_{23}s_{13}e^{i\delta}
+c_{12}^2c_{23}^2s_{13}^2e^{2i\delta}|\,,\label{eq15}
\end{align}
while in the {\it IH} case it is as follows
\begin{equation}
s_{23}^2=c_{23}^2\,.\label{eq16}
\end{equation}
Equation~(\ref{eq15}) permits one to determine the
\textit{CP}-violating phase $\delta$, while Eq.~(\ref{eq16})
determines the angle $\theta_{23}$. With using the data from
Eqs.~(\ref{eq4}) we can obtain that $\delta\approx 100$\!\r{} in
the NH case, in contrast to the frequently used value of
$\delta=0$\!\r{}. However, the obtained value of $\delta$ is
different also from the value $\delta=180$\!\r{} \cite{Fogli2012}.
For this reason we consider that Eq.~(\ref{eq15}) is not
fulfilled, that is the condition of the $\mu-\tau$ symmetry
results in preferability of the IH case of the neutrino mass
spectrum. In this case, it is possible to estimate the values of
all the neutrino mass observables $m_{a}$, $m_{\beta }$ and
$m_{\beta\beta}$, if the minimal neutrino mass is of the order of
$\lambda$. In~eV, they are as follows
\begin{equation}
m_{a}\approx0.034,\,m_{\beta}\approx0.049,\,
m_{\beta\beta}\approx0.00005\,.\label{eq17}
\end{equation}
The dependences of $m_{\beta\beta}$ versus the minimal neutrino
mass $m_{0}$ in the model of bimodal neutrinos are shown in
Fig.~\ref{Fig:5}, in the range of $m_0$ between $0$ and $1$~eV.
The characteristic feature of these results is the absence of the
bottom limitation for $m_{\beta\beta}$ in both NH and IH cases,
because in the bimodal neutrinos case $m_{\beta\beta}$ depends
only on just one term involving $m_0$, that is $m_1$ or $m_3$ in
the NH or IH case, respectively.

\section{Conclusion}
\label{Section6} %
The properties of the neutrino are rather mysterious, and both
intensive theoretical and experimental studies are necessary for
determination of the nature and the characteristics of these
elementary particles. The construction and development of adequate
phenomenological models of neutrino, which generalize the SM in
the neutrino sector is one of the ways to interpret and predict
the experimental results, as well as to develop the future GUT. At
present, one of the first-priority problem in both the theoretical
and the experimental investigations is the ascertainment of the
neutrino type, that is the Dirac type or the Majorana type. It may
occur rather unexpected that, possibly, this problem has no
unambiguous solution and the neutrino can be bimodal, as was
considered in the present paper. The other important problems,
which should be solved in the process of the theoretical and
experimental investigations are the determination of the absolute
mass scale, the characteristics of the {\it CP} violation of the
neutrino, and the different correlations between the numerous
neutrino parameters.

In the present paper, the possible values of the neutrino mass
observables $m_{a}$, $m_{\beta}$ and $m_{\beta\beta}$ were
calculated on the basis of the most recent experimental data. It
was found that the minimal mass neutrino range, where
$m_{\beta\beta}$ vanishes is divided into two sub-ranges with
limitation from the bottom between them for the $m_{\beta\beta}$
values in the case of the normal hierarchy. It takes place both
for the neutrino mass spectrum in the process of neutrinoless
double-beta decay with {\it CP} violation, and even in the
intermediate range of $m_0$ for the decays with {\it CP}
conservation. For the investigation of the neutrino properties,
the $3+2+1$ phenomenological neutrino model with three active
neutrinos and three sterile neutrinos was proposed. The model
permits reducing the number of sterile neutrinos, if the
model-independent experimental restrictions on their number will
be established. In the framework of this model and with allowance
for the recent experimental data, the values of the neutrino mass
observables $m_{a}$, $m_{\beta}$ and $m_{\beta\beta}$ were
obtained, and also the estimations of masses $m_{si}\equiv M_i$ of
the light sterile neutrinos [see Eqs.~(\ref{eq14})] were made. As
is known, for experimental determination of the observables
$m_{a}$, $m_{\beta}$ and $m_{\beta\beta}$, numerous experiments
are currently carried out and planned, namely, the experiments on
search the neutrinoless double-beta decay, on determination of the
form of the tritium $\beta$ decay spectrum, as well as the
cosmological observations. The theoretical estimations of the
values of $m_{a}$, $m_{\beta}$, $m_{\beta\beta}$ and $m_{si}$
given above in this paper can be used for the interpretation and
prediction of the results of these experiments.

\begin{acknowledgments}
Authors are grateful to S.~V.~Semenov, Yu.~S.~Luto\-stansky,
D.~K.~ Nadezhin and I.~V.~Panov for useful discussions and
remarks. This work was supported by the Russian Foundation for
Basic Research (Grant No. 11-02-00882-a).
\end{acknowledgments}


\section*{References}
\begin{thebibliography}{10}

\bibitem{Fogli2012} G. L. Fogli, E. Lisi, A. Marrone, D. Montanino,
A. Palazzo, and A. M. Rotunno, arXiv:1205.5254. 

\bibitem{STV2011}  T. Schwetz, M. T´ortola, and J. W. F. Valle,
arXiv:1103.0734. 

\bibitem{GGN2003} M. C. Gonzalez-Garcia and Y. Nir, Rev. Mod. Phys. \textbf{75},
345 (2003). 

\bibitem{BGFJ2012} G. C. Branco, R. Gonzalez Felipe, and F. R. Joaquim, Rev. Mod.
Phys. \textbf{84}, 515 (2012). 

\bibitem{Dirac} P. A. M. Dirac, Proc. Roy. Soc. A \textbf{117}, 610 (1928);
{\it ibid.} \textbf{118}, 351 (1928). 

\bibitem{Majorana} E. Majorana, Nuovo Cimento \textbf{14}, 171 (1937). 

\bibitem{Wolfenstein1981} L. Wolfenstein, Nucl. Phys. \textbf{B186}, 147
(1981).

\bibitem{Petcov1982} S. T. Petcov, Phys. Let. B \textbf{110}, 245
(1982).

\bibitem{Doi1983} M. Doi, M. Kenmoku, T. Kotani, H. Nishiura, and
E. Takasugi, Prog. Theor. Phys. \textbf{70}, 1331 (1983).

\bibitem{Valle1983} J. W. F. Valle, Phys. Rev. D \textbf{27}, 1672
(1983).

\bibitem{ADM2011} R. Allahverdi, B. Dutta, and R. N. Mohapatra, Phys. Lett. B
\textbf{695}, 181 (2011). 

\bibitem{MP2011} A. C. B. Machado and V. Pleitez, Phys. Lett. B \textbf{698},
128 (2011). 

\bibitem{CL2011} C. S. Chen and C. M. Lin, arXiv:1101.4362. 

\bibitem{MS2006} R. N. Mohapatra and A. Y. Smirnov, arXiv:hep-ph/0603118.

\bibitem{Fritzsch2009} H. Fritzsch, arXiv:0902.2817. 

\bibitem{Gaponov2011} Yu. V. Gaponov,  Phys. At. Nuclei \textbf{74}, 272
(2011). 

\bibitem{Khruschov2011} V. V. Khruschov, arXiv:1106.5580; Preprint IAE-6692/2, Moscow
(2012). 

\bibitem{Abazajian2012} K. N. Abazajian {\it et al}., arXiv:1204.5379. 

\bibitem{Komatsu2011} E. Komatsu {\it et al}. [WMAP Collaboration], Astrophys. J. Suppl.
\textbf{192}, 18 (2011). 

\bibitem{Nakamura2010} K. Nakamura {\it et al}. [PDG], J. Phys. G \textbf{37}, 075021
(2010). 

\bibitem{BGG1999} S. M. Bilenky, C. Giunti, and W. Grimus, Prog. Part. Nucl.
Phys. \textbf{43}, 1 (1999).

\bibitem{OW2008} E. W. Otten and C. Weinheimer, Rep. Prog. Phys. \textbf{71},
086201 (2008). 

\bibitem{KK2001} H. V. Klapdor-Kleingrothaus {\it et al}.,
Nucl. Phys. Proc. Suppl. \textbf{100}, 309(2001). 

\bibitem{Aalseth2004} C. Aalseth {\it et al}., arXiv:hep-ph/0412300. 

\bibitem{FY1986} M. Fukugita and T. Yanagida, Phys. Lett. B \textbf{174}, 45
(1986). 

\bibitem{Mueller2011} Th. A. Mueller {\it et al}., Phys. Rev. C \textbf{83}, 054615
(2011).

\bibitem{Abdurashitov2006} J. N. Abdurashitov {\it et al}., Phys. Rev. C \textbf{73}, 045805
(2006). 

\bibitem{GL2010} C. Giunti and M. Laveder, arXiv:1006.3244. 

\bibitem{Palazzo2012} A. Palazzo, arXiv:1201.4280. 

\bibitem{Aguilar2001} A. Aguilar {\it et al}. [LSND Collaboration], Phys. Rev. D
\textbf{64}, 112007 (2001). 

\bibitem{AguilarArevalo2010} A. A. Aguilar-Arevalo {\it et al}. [MiniBooNE Collaboration], Phys. Rev.
Lett. \textbf{105}, 181801 (2010). 

\bibitem{BG2012} S. M. Bilenky and C. Giunti, arXiv:1203.5250. 

\bibitem{Fritzsch2011} H. Fritzsch, Z.-z. Xing, and S. Zhou, arXiv:1108.4534. 

\bibitem{MB2012} D. Meloni and G. Blankenburg, arXiv:1204.2706. 

\bibitem{MR2006} A. Merle and W. Rodejohann, Phys. Rev. D \textbf{73}, 073012
(2006).

\bibitem{CPP2006} B. C. Chauhan, J. Pulido, and M. Picariello, Phys. Rev. D \textbf{73},
053003 (2006). 

\bibitem{PP2004} S. Pascoli and S. T. Petcov, Phys. Lett. B \textbf{544}, 239
(2002); {\it ibid.} \textbf{580}, 280 (2004). 

\bibitem{PPS2006} S. Pascoli, S. T. Petcov, and T. Schwetz, Nucl. Phys. \textbf{B734}, 24
(2006). 

\bibitem{GL2012} W. Grimis and L. Lavoura, arXiv:1207.1678. 

\end{thebibliography}
\end{document}